\documentclass[galaxies,article,accept,moreauthors,pdftex,10pt,a4paper]{mdpi} 
\firstpage{1} 
\articlenumber{x}
\doinum{10.3390/------}
\pubvolume{xx}
\pubyear{2017}
\copyrightyear{2017}
\externaleditor{Academic Editor: name}
\history{Received: date; Accepted: date; Published: date}

\pdfoutput=1

\Title{The University of Michigan Centimeter-Band All Stokes Blazar Monitoring Program: Single-dish Polarimetry as a Probe of Parsec-scale Magnetic Fields}

\Author{Margo F. Aller $^{1}$\orcidA{}, Hugh D. Aller $^{1}$ and Philip A. Hughes$^{1,}$}

\AuthorNames{Margo Aller, Hugh Aller, and Philip Hughes}

\address{%
$^{1}$ \quad Department of Astronomy, University of Michigan, Ann Arbor MI 48109-1107 USA 1; mfa@umich.edu}

\corres{Correspondence: mfa@umich.edu; Tel.: 001-734-764-3465}

\abstract{The University of Michigan 26-m paraboloid was dedicated to obtaining linear polarization and total flux density observations of blazars from the mid-1960s until June 2012 providing an unprecedented record tracking centimeter-band variability over decades at 14.5, 8.0, and 4.8 GHz for both targeted objects and members of flux-limited samples. In the mid-1970s through the mid-1980s, and during the last decade of the program, observations were additionally obtained of circular polarization for a small sample of radio-bright (S>5Jy), active sources. Key program results include evidence supporting class-dependent differences in the magnetic field geometry of BL Lac and QSO jets, identification of linear polarization changes temporally associated with flux outbursts supporting a shock-in-jet scenario, and determination of the spectral evolution of the Stokes V amplitude and  polarity for testing proposed models. Recent radiative transfer modeling during large flares supports a jet scenario with a  kinetically-dominated, relativistic flow at parsec scales with embedded turbulent magnetic fields and dynamically-weak ordered components which may be helical; the circular polarization observations are consistent with  linear-to-circular mode conversion within this turbulent jet environment.}


\begin{document}


\section{Introduction}

The University of Michigan Radio Astronomy Observatory’s 26-m paraboloid (UMRAO) was dedicated to monitoring the polarization of blazar jets from the mid-1960s following the first documented discovery of centimeter-band total flux density variability with this instrument \cite{DEN65-Science} and the subsequent first report of linear polarization variability \cite{ALL67-ApJ}. The approximately one million observations obtained until the closure of the facility in June 2012 explored the temporal and spectral properties of the radio-band linear polarization variability and provided the basis for an interpretation based on a shock-in-jet paradigm, commonly invoked to explain the variability of the total flux density and linear polarization since the mid-1980s. During the last decade of operation a sample of about a dozen blazars (selected during initial observing runs from a larger target list of 35 Active Galactic Nuclei [AGNs] with known or suspected centimeter-band circular polarization) was monitored to investigate the variability properties of the circular polarization on decadal timescales, potentially providing a link between the jet and the central engine. While not discussed here, the UMRAO data have been invoked in countless broad band studies of flaring blazars and to calibrate and to interpret the more sparsely-sampled  Very Long Baseline Array (VLBA) imaging data.  Currently the archival data are being used to constrain radiative transfer models incorporating propagating shocks in order to determine key shock and jet parameters characterizing the flows, and their evolution with time. These jet parameters include the geometry of the magnetic field at parsec scales. We summarize here key polarization results from the program based on past and ongoing studies using the UMRAO data.

\begin{figure}[H]
\centering
\includegraphics[width=3.8in]{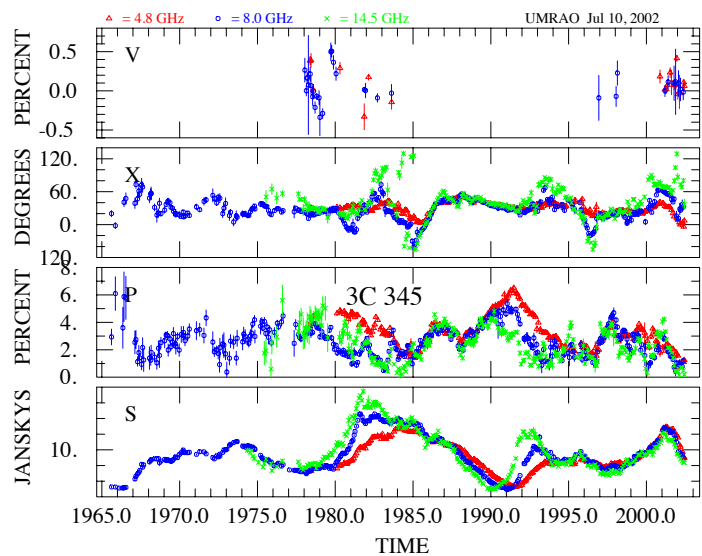}
\caption{ From bottom to top: Long-term total flux density (S), fractional linear polarization (P\%),  electric vector position angle (EVPA or $\chi$), and fractional circular polarization (V\%) for the quasar 3C 345. The three centimeter-band frequencies are symbol and color coded as denoted at the top left. Polarity changes in V occurred at both 4.8 and 8.0 GHz in 1978 - 1984 contemporaneous with a large self-absorbed  outburst in total flux density.}
\end{figure}   

\section{The Data}
Figure 1 shows the time variability of the total flux density and the fractional linear and circular polarization and their spectral evolution for the bright quasar 3C~345.  Although initially UMRAO data were obtained at 8 GHz only, the development of radiometer-polarimeter systems to detect linear polarization  at 14.5 GHz in the 1970s, and shortly thereafter both linear and  circular polarization at 4.8 GHz, added spectral information. Program sources included both ad hoc sources selected on the basis of flaring or other characteristics making them of special interest, as well as flux-limited samples used to investigate class properties; the latter include the UMRAO BL Lac sample \cite{ALL99-ApJ}, the Pearson Redhead sample \cite{ALL92-ApJ, ALL03-ApJ}, and the Stangellini sample of GHz-peaked sources \cite{ALL02-Proceedings}.  The observing cadence for each source was based on the time scale of the total flux and linear polarization variability using the archived data as a guide and varied from two times a week for the most variable sources at the highest UMRAO frequency, 14.5 GHz, where the variations are highest in amplitude and most rapid, to only a few times a year for the less variable members of the flux-limited samples. Typically 24 program sources were observed in a daily linear polarization run, and observations of calibrator sources were interleaved at 1.5 to 3 hour intervals in order to monitor pointing and  gain changes, and to verify the amplitude of the instrumental polarization. Circularly polarized emission (Stokes V) is typically only tens of mJy even in the stronger sources, and with the small aperture of the Michigan dish, Stokes V measurements required daily source integration times of approximately an hour, and averaging of these data over many days to obtain a 3-$\sigma$ detection. Generally circular polarization runs at 4.8 and 8 GHz were conducted over 2 consecutive days, while at 14.5 GHz where tropospheric effects were more severe, runs were 3 consecutive days in duration.  The circular polarization data shown in this paper are two-week or monthly averages, and  due to this binning, required to obtain significant detection levels of this weak emission, variations on timescales of a week or less cannot be studied. Comparisons of the UMRAO observations with contemporaneous centimeter-band circular polarization measurements obtained with the Effelsberg dish \cite{MYS17-AA} and with the VLBA \cite{HOM09-ApJ} show consistent amplitudes and polarity, and these complementary data confirm the accuracy of these difficult measurements. 

 \section{Characteristics of the Linear Polarization}

   An ongoing goal of the program is to use the fractional linear polarization and the EVPA to investigate the geometry and degree of order of the magnetic field in the emitting region. The amplitude of the fractional linear polarization during relatively quiescent time windows gives information on the degree of order of the magnetic field in the underlying quiescent jet. As shown in Figure 1, these  are typically only of order a few percent during flaring and they  range up to about 15\%  during flaring in a handful of sources at 14.5 GHz. As discussed in \cite{HUG05-ApJ}, the low values are consistent with the presence of a turbulent or tangled magnetic field in the quiescent jet. The UMRAO data formed the basis for the shock-in-jet interpretation of the variability in the 1980s whereby the variability during large outbursts is attributed to propagating shocks which compress the tangled magnetic field during their passage, increasing the degree of order, and producing outbursts in polarized flux temporally associated with the total flux density flaring.

\begin{figure}[H]
\centering
\includegraphics[width=3.8in]{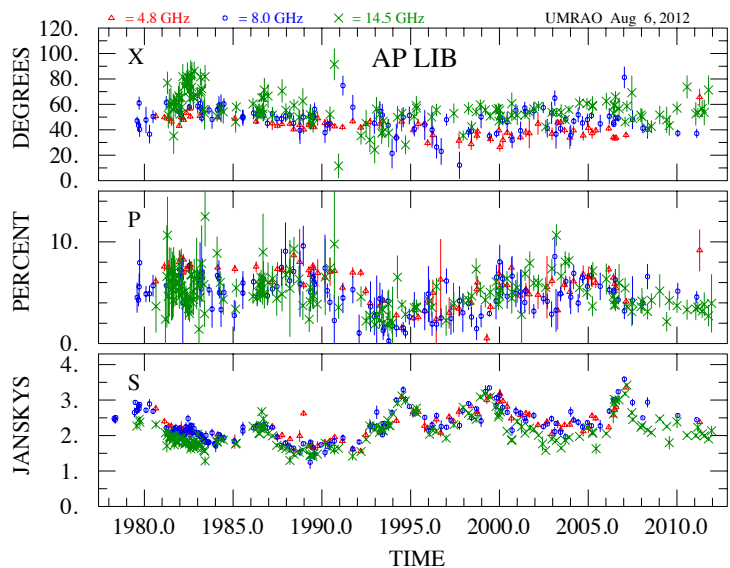}
\caption{From bottom to top: Long-term total flux density (S), fractional linear polarization (P\%), and EVPA for AP Librae. Two-week-averaged data are shown. No rotation measure was determined for this source in \cite{HOV12-AJ}: the fractional polarization was weak (0.7\%) during the MOJAVE epoch analyzed, April 28, 2006.}
\end{figure}  

\subsection{Statistical Results on Maximum Amplitude, and Preferred EVPA Orientation by Optical Class}

The EVPA is orthogonal to the magnetic field direction for the special case of a transparent non-relativistic flow.  Faraday rotation can produce a shift of the observed EVPA from its intrinsic value, but a correction can be made using rotation measures determined from  Very Large Array (VLA) or VLBA measurements, e.g. \cite{HOV12-AJ}. At 14.5 GHz the shift due to Faraday rotation is generally very small, and only a few sources in the UMRAO program show the characteristic signature of Faraday rotation, a $\lambda^{2}$ separation, in the 14.5 to 4.8 GHz EVPA light curves. The difference between the most common EVPA orientation at 4.8 GHz and the VLBI structural axis, $\theta$, supported a class-dependent picture between the BL Lacs and the QSOs. For the BL Lacs the data commonly showed large-amplitude changes in fractional polarization (polarized flares) and rapid variations in EVPA, while the distribution of  the difference between the intrinsic EVPA and $\theta$ exhibited a flat distribution. This result suggested that shocks with a range of obliquities are prevalent in BL Lac flows, and comparison with the VLBA data available at the time suggested that the integrated polarization is dominated by emission from the unresolved cores \cite{ALL99-ApJ}. For the QSOs in the Pearson-Readhead sample the  intrinsic EVPA-$\theta$ difference was sharply-peaked near 90$^{\circ}$ and polarization changes were slower \cite{ALL92-ApJ,ALL03-ApJ}. This behavior suggested that the polarization in these sources is dominated by the contribution from the underlying quiescent jet where shear imposes a modest degree of order on the tangled magnetic field. 
    
\subsection{Long-term Preferred EVPAs: An unusual BL Lac object}
 Some sources show a long-term preferred EVPA extending over many flares in total flux density consistent with a long-term memory. An example source is the flat spectrum, core-dominated, low-spectral-peaked BL Lac object AP Librae (PKS 1514-24). See Figure 2. However, the presence of a multi-decade preferred EVPA is unusual for a BL Lac object. As can be seen in MOJAVE polarization maps, E is nearly perpendicular to the local flow direction in the jet of this source in many epochs. Possible scenarios to explain this behavior, more typical of QSOs, are that this flow is sheared, or that the magnetic field is dominated by a helical component and not the turbulent one, or possibly  that oblique shocks are present in a flow with a turbulent magnetic field. 

\begin{figure}[H]
   \begin{center}$
\begin{array}{cc}
   \includegraphics[width =2.9in]{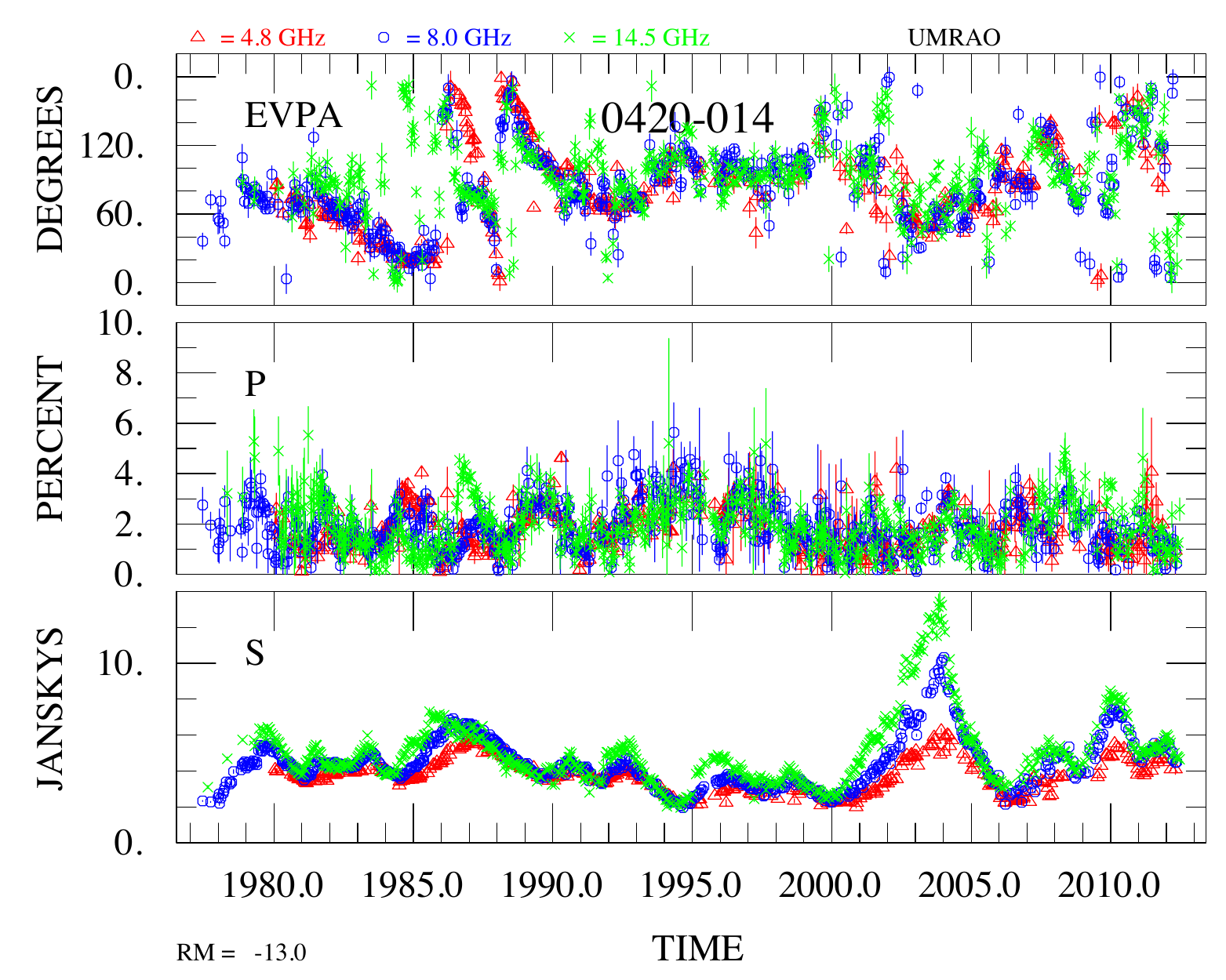} &
\hfill
  \includegraphics[width=2.9in]{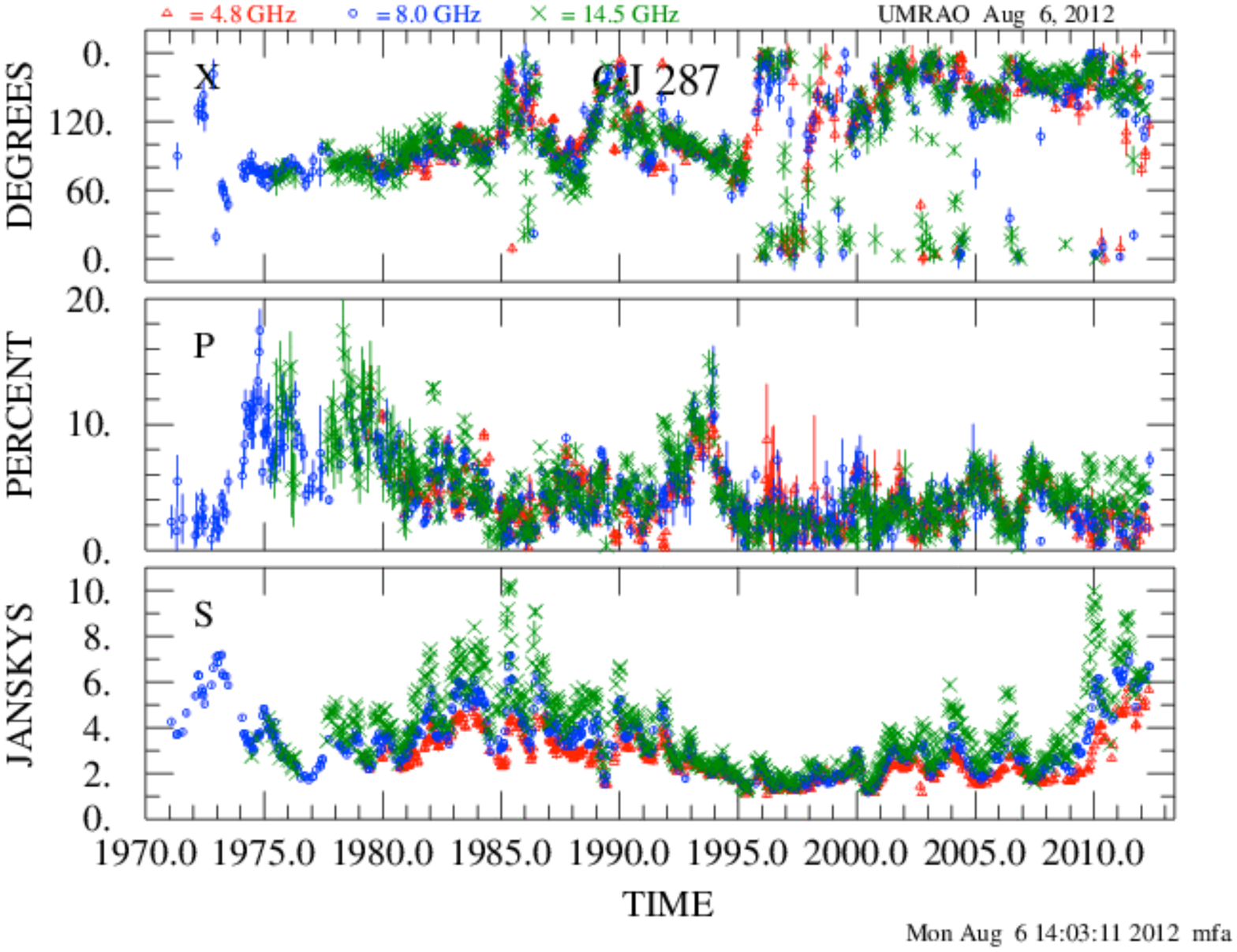}
\end{array}$
\end{center}
\caption{Example of EVPA rotations in the quasar 0420-014 (left) and in the BL Lac object OJ 287 (right). Due to n$\pi$ ambiguities in the EVPA determination, the range is restricted to  180$^{\circ}$ in these plots.}
\end{figure}  

\subsection{Long-term Ordered Variations in EVPA: Rotations}
 Monotonic changes in the EVPA over tens of degrees have been detected in both radio and optical polarimetry data. These ordered changes were noted in the early UMRAO data, e.g. \cite{ALL81-ApJ} and first attributed to a rotating structure in the radio-emitting region but subsequently explained in terms of a random walk generated by the evolution of a turbulent magnetic field \cite{JON85-ApJ}. Examples are shown in Figure 3 illustrating the range of the monotonic ordered change in the radio-band and its timescale. Recent analyses based on polarimetry data in the optical band, have suggested that both deterministic origins such as a disturbance moving in a helical jet, with the EVPA rotation occurring either upstream \cite{MAR10-ApJ} or downstream \cite{ANG16-MNRAS} of the shock, as well as stochastic process \cite{KIE16-AA,KIE17-MNRAS} may be present. The properties of the ordered changes in the two bands have both similarities and  differences as summarized in Table 1, and may require different physical explanations. Of particular note are the timescales and the location of the region where the rotation is believed to occur (further upstream for the optical cases, where the jet flow may be dominated by a ordered, helical magnetic field). Interpretation of these ordered changes depends on whether the flow in the emitting region is magnetically or kinetically dominated, and both scenarios have been a topic of recent investigation.  
While we have preferred the stochastic explanation for the EVPA rotations apparent in the UMRAO data, M. Cohen has been able to explain the  UMRAO time series for OJ 287 in terms of an MHD model (presentation at this conference), so interpretation of this phenomenon in both bands is ambiguous.

 \begin{table}[H]
\caption{Observed and Inferred Polarization Properties During Large EVPA Rotation Events}
\begin{tabular}{ccc}
\toprule
\textbf{Property}	& \textbf{Optical band}	   & \textbf{Radio band}\\
\midrule
Timescale		& A week to a few months & A few years \\
Range of $\Delta$EVPA	& >90$^{\circ}$ to 720$^{\circ}$	 &  A few hundred degrees \\
P\% during Rotation    & Mean P\% $<$ P\% during non-event    &  P\% over a range (see Fig. 3)  \\
No. cells/regions &  Many                                  & A few VLBI-scale components \\
Emission Site          & Shock/moving disturbance in helical jet     &  Disturbance in turbulent kinetic or MHD jet \\
Emission Location       & Upstream of the 43 GHz core & Parsec-scale jet  \\
\bottomrule
\end{tabular}
\end{table}

\section{Ordered Changes During Flaring: The Shock Scenario and Radiative Transfer Modeling}
    Large radio-band outbursts in total flux are envelopes over individual flares which are commonly blended in centimeter-band data with durations of months to years. Often the contributions from the individual flares are better resolved in temporally-associated outbursts in linear polarization. Otherwise the contributions to the emission by the individual flares can  be  separated using a combination of the structure in the light curves and the theoretically-expected outburst shape \cite{HUG11-ApJ}.  During the outbursts  an ordered change in the fractional linear polarization with an associated swing in EVPA through tens of degrees can often be seen in the UMRAO data. Examples are shown in \cite{ALL14-ApJ}. These changes are consistent with the signature expected from the passage of a shock which compresses the tangled magnetic field in the quiescent jet and produces an increase in the emissivity and the degree of order of the magnetic field. 

 In order to obtain information on several jet and shock parameters which are not directly observed, we are currently using the UMRAO data to constrain simulations incorporating propagating shocks oriented at an arbitrary angle to the flow direction indicated by complementary VLBA observations. The shocks compress the magnetic field which is assumed to be predominantly turbulent before the passage of the shocks. Each model has only 6 key free parameters, and it is very well constrained by the spectral evolution of both the linear polarization and the total flux density. This ongoing work has been described in a series of papers which have shown that the model can reproduce the main features in the data during several strong outbursts in the blazars 0420-014, 0716+714, OJ 287, 1156+295, and OT 081 \cite{ALL14-ApJ,HUG15-ApJ,ALL16-GAL}. The modeling is currently being used to study the evolution of the flow on decadal times scales and to constrain the effect of a helical magnetic field on the linear polarization light curves. Both the amplitude of the fractional linear polarization and its spectral evolution can be reproduced with the model, and the polarization is a sensitive test of the shock scenario. The details of the EVPA evolution are not well reproduced, however, and this may be due to simplifications of the model. An important result is that the predicted large (tens of degrees) frequency-dependent separation  of the EVPA light curves with a spread by as much as $\Delta$EVPA=35$^{\circ}$ for the case of a strong helical magnetic field considered in \cite{ALL16-GAL} is inconsistent with the behavior seen in the UMRAO data.

\section{Circular Polarization: a potential probe of the large scale  helical and longitudinal} magnetic field
Observations of circular polarization were carried out in the late 1970s and early 1980s at 4.8 and 8.0 GHz. These measurements were subsequently dropped from the Michigan program because of the long integration times required for detections and the limited number of sources for which a 3-$\sigma$ detection was expected. However, they were resumed in 2002 with the addition of a new radiometer-polarimeter system operating at 14.5 GHz, thereby increasing the spectral coverage. The new program was motivated in part by the suggestion by Ensslin \cite{ENS03-AA} that the sign of Stokes V measures the rotational direction of the central engine, and that polarity reversals would be rare, providing a direct observational test of the Ensslin model by examining the long-term stability of the polarity. This new UMRAO program differs in scope from other centimeter-band programs which were concerned with detection statistics, or were conducted at one frequency only providing no information on the bandwidth of the emission, and generally time windows include only a few years e.g. \cite{KOM84-MNRAS,HOM06-AJ}. Specific observational goals of the UMRAO program were to delineate the amplitude range of the Stokes V variations, to look for relations between the 4 Stokes parameters as tests of proposed emission mechanisms, and to investigate whether the polarity of Stokes V is stable on decadal timescales.

\begin{figure}[H]
   \begin{center}$
\begin{array}{cc}
   \includegraphics[width =2.9in, bb = 54 55 712 568,clip]{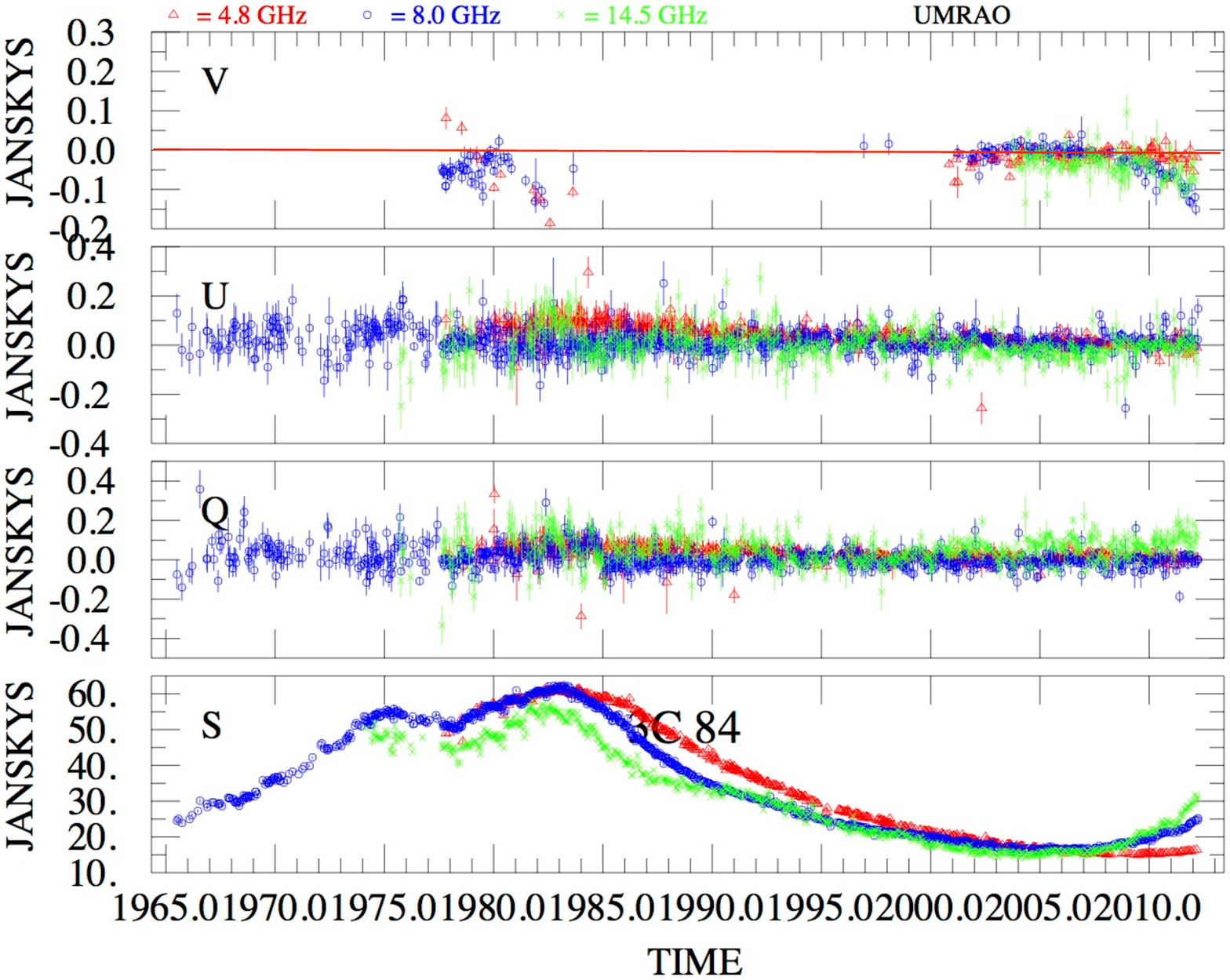} &
\hfill  
  \includegraphics[width=2.9in, bb = 54 55 712 568,clip]{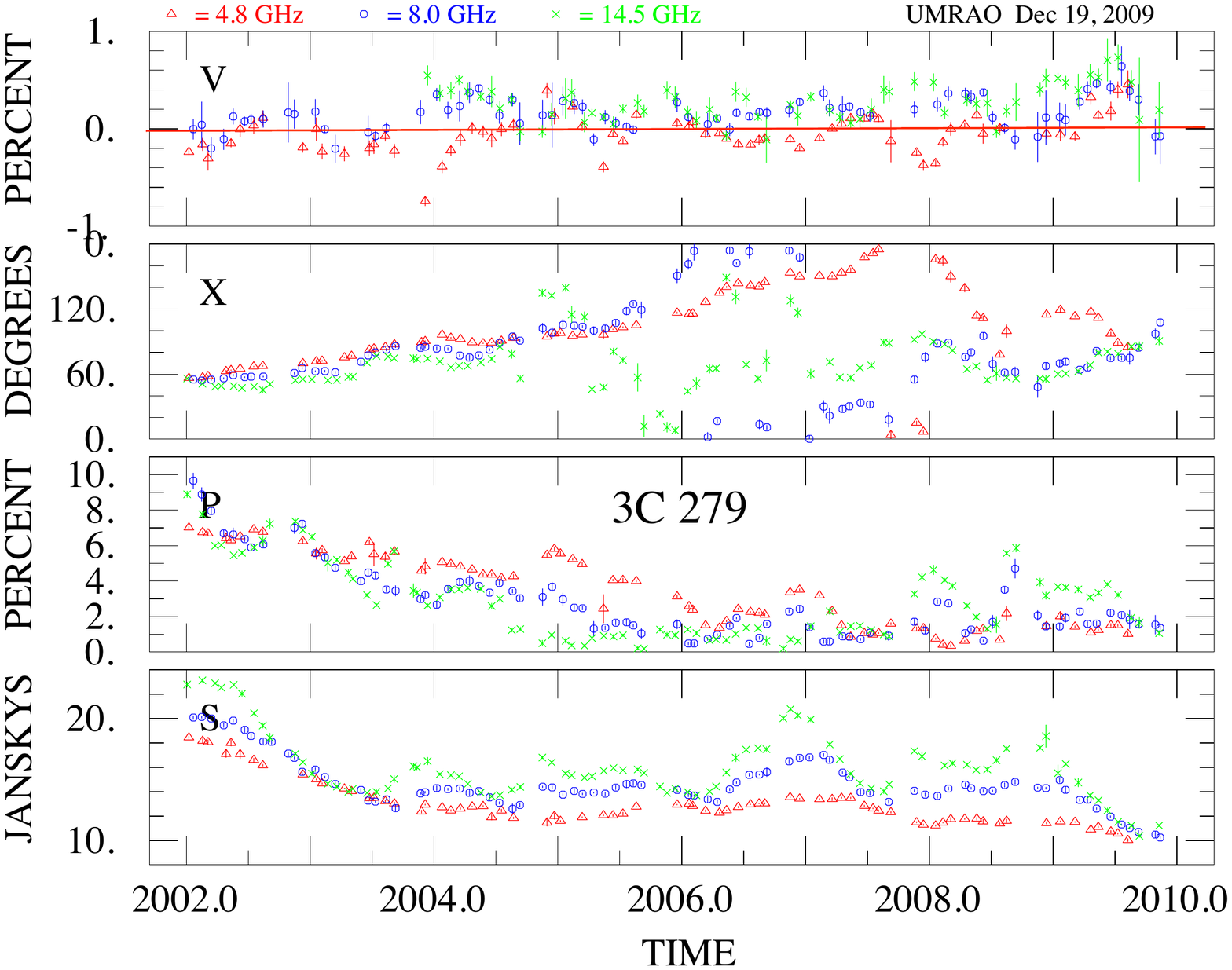}
\end{array}$
\end{center}

\caption{Left: bottom to top: Long-term 2-week averaged Stokes parameters for the radio galaxy 3C~84 showing polarity changes at 4.8 GHz during the late-1970s through the early-1980s. Note the change in amplitude of V at 14.5 GHz  associated with the commencement of new flaring circa 2008.  Right: total flux density, fractional linear polarization, EVPA and fractional circular polarization for 3C 279 illustrating polarity reversals which are most evident at 4.8 GHz.}
\end{figure}  

 Figure 4 shows circular polarization results for two of the best-sampled program sources, 3C~84 and 3C~279. For 3C~84 the fractional linear polarization was <1\% until the start of new activity in 2008; because of the very low fractional linear polarization amplitude resulting in low signal-to noise in the polarization light curve, the data for this source are shown in the form of Stokes parameters.  The total flux density light curve (Stokes I: lower panel) for 3C 84  exhibits only a few long-term outbursts attaining an usually-high peak amplitude near 60 Jy. The upper panel shows polarity reversals at 8 and 4.8 GHz, while at 14.5 GHz the polarity remained constant (negative). Note that while the fractional circular polarization is less than 0.4\%, the circularly polarized emission is, in fact, relatively strong, even during times when the amplitude of the linear fractional polarization is very low. While there are clear cases of positive polarity, the polarity is mainly negative.  3C 279 (right) has shown polarity reversals at 8 and 4.8 GHz and frequency-dependent differences in sign at a common epoch.

With the exception of 3C~84, there is no clear correlation between the amplitude of the variability in fractional circular polarization and that in total flux or fractional linear polarization. The polarity exhibits a variety of behavior patterns.
OV-236 is the only UMRAO program source which exhibits a constant polarity over decades of measurements during a time window with several major outbursts. In spite of the polarity reversals noted above, there are preferred polarities in a given source, lasting over many outbursts and  consistent with the presence of a long-term memory. The polarity reversals sometimes occur at times at which there is evidence for self-absorption in the total flux density and linear polarization multi-frequency light curves. In these events, the sign changes may be due to opacity effects whereby different segments of the jet dominate the source-integrated emission. No reversals were observed at 14.5 GHz in the UMRAO data, but these reversals might have been masked by the lower signal-to-noise in the data at this frequency, as reversals have been commonly found from Stokes V monitoring at millimeter wavelengths \cite{AGU17-GAL}. The accumulated data have continued to support a model for the production of circular polarization by linear-to-circular mode conversion as proposed in earlier work \cite{ALL03-ApSS}.

\section{Summary and Conclusion}
The longterm, multifrequency polarization observations obtained by the UMRAO program continue to be important for increasing our understanding of the physical conditions in the parsec scale regions of blazar jets. The linear polarization is currently being used to constrain shock models in order to determine intrinsic shock and jet flow parameters, and several flares in both QSOs and BL Lac objects have now been successfully modeled supporting the validity of the adopted shock-in-jet model, at least in the case of strong flares. 

While observations of circular polarization in principle can provide additional constraints on any large scale helical and longitudinal magnetic field components and the composition of the jet, both the measurements of this emission and the calibration of the data are difficult, and the time variation of the polarity exhibits a variety of behaviors and not a simple one as expected from the simple Ensslin model \cite{ENS03-AA}; preferred polarities are present, consistent with the presence of a longterm memory, but polarity reversals are also common, which are not always associated with opacity changes as indicated by the spectral evolution apparent in the light curves. 
The UMRAO circular polarization data were unable to identify unambiguous signatures of an underlying unidirectional magnetic field. One might expect to see these signatures more clearly with millimeter data which probe emission upstream, but the data reported so far from the program Polarimetric Monitoring of Active Galactic Nuclei at Millimeter Wavelengths (POLAMI) \cite{AGU17-GAL} identify similar maximum amplitudes of only a few percent and polarity changes consistent with a turbulent magnetic field picture. As a further complication in the interpretation, new theoretical work incorporating helical magnetic fields has shown that polarity reversals can be attributed to the growth of instabilities in the flow \cite{DUT17-PROC}, and these might contribute to the masking of a jet-central engine connection associated with an underlying large scale magnetic field. While recent simulations reproducing the sign and amplitude of Stokes V assuming a turbulent jet also support the turbulent jet scenario \cite{MAC17-ApJ} with generation of the emission by Faraday conversion, important questions remain about the geometry and role of a large scale magnetic field at parsec scales and the feasibility of unambiguously identifying a jet-central engine connection from the centimeter-band polarimetry data.

\acknowledgments{This work was funded in part by a series of grants from the NSF (most recently AST-0607523) and by a series of Fermi G.I. awards from NASA (NNX09AU16G, NNX10AP16G, NNX11AO13G, and NNX13AP18G). Funds for operation of UMRAO were provided by the University of Michigan.  M.A. received travel support from RadioNet. RadioNet has received funding from the European Union’s Horizon 2020 research and innovation program under grant agreement No. 730562.}

\authorcontributions{Hugh D. Aller and Margo F,  Aller obtained, reduced, and analyzed the observations. Philip A. Hughes led the radiative transfer modeling program and developed the computer codes used to simulate the light curves. All participated in the analysis of the data described here.}

\conflictsofinterest{The authors declare no conflict of interest.}





\end{document}